# Multiple Attractor Cellular Automata (MACA) for Addressing Major Problems in Bioinformatics

[1] Pokkuluri.Kiran Sree, [2] Inampudi Ramesh Babu & [3] SSSN Usha Devi Nedunuri

[1] Research Scholar, Dept of CSE, Jawaharlal Nehru Technological University, Hyderabad

[2] Professor, Dept of CSE, Acharya Nagarjuna University, Guntur

[3] Assistant Professor, Dept of CSE, JNTUK, Kakinada

profkiransree@gmail.com, drirameshbabu@gmail.com, usha.jntuk@gmail.com

*Abstract*

CA has grown as potential classifier for addressing major problems in bioinformatics. Lot of bioinformatics problems like predicting the protein coding region, finding the promoter region, predicting the structure of protein and many other problems in bioinformatics can be addressed through Cellular Automata. Even though there are some prediction techniques addressing these problems, the approximate accuracy level is very less. An automated procedure was proposed with MACA (Multiple Attractor Cellular Automata) which can address all these problems. The genetic algorithm is also used to find rules with good fitness values. Extensive experiments are conducted for reporting the accuracy of the proposed tool. The average accuracy of MACA when tested with ENCODE, BG570, HMR195, Fickett and Tongue, ASP67 datasets is 78%.

## Introduction

Many interesting problems in bioinformatics can be addressed by Cellular Automata Classifier. Predicting the structure of protein with the topology of the chain, finding the protein coding region and finding the promoter region can be solved very easily with CA. The tree dimensional arrangement of amino acid sequences can be described by tertiary structure. They can be predicted independent of each other. Functionality of the protein can be affected by the tertiary structure, topology and the tertiary structure. Structure aids in the identification of membrane proteins, location of binding sites and identification of homologous proteins (Debasis Mitra, 2004) to list a few of the benefits, and thus highlighting the importance, of knowing this level of structure  This is the reason why considerable efforts have been devoted in predicting the structure only. Knowing the structure of a protein is extremely important and can also greatly enhance the accuracy of tertiary structure prediction. Furthermore, proteins can be classified according to their structural elements, specifically their alpha helix and beta sheet content.

The primary goal of bioinformatics is to increase the understanding of biological processes. What sets it apart from other approaches, however, is its focus on developing and applying computationally intensive techniques to achieve this goal. Examples include: pattern recognition, data mining, machine learning algorithms, and visualization. Major research efforts in the field include sequence alignment, gene finding, genome assembly, drug design, drug discovery, protein structure alignment, protein structure prediction, prediction of gene expression and protein–protein interactions, genome-wide association studies, and the modeling of evolution.

Protein coding sequences are often considered to be basic parts, in fact proteins coding sequences can themselves be composed of one or more regions, called protein domains. Thus, a protein coding sequence could either be entered as a basic part or as a composite part of two or more protein domains.

1.      The N-terminal domain of a protein coding sequence (Eric E. Snyder, 2002) is special in a number of ways. First, it always contains a start codon, spaced at an appropriate distance from a ribosomal binding site. Second, many coding regions have special features at the N terminus, such as protein export tags and lipoprotein cleavage and attachment tags. These occur at the beginning of a coding region, and therefore are termed Head domains.

2.      A protein domain is a sequence of amino acids which fold relatively independently and which are evolutionarily shuffled as a unit among different protein coding regions. The DNA sequence of such domains must maintain in-frame translation, and thus is a multiple of three bases. Since these protein domains are within a protein coding sequence, they are called Internal domains. Certain Internal domains have particular functions in protein cleavage or splicing and are termed Special Internal domains.

3.      Similarly, the C-terminal domain of a protein is special, containing at least a stop codon. Other special features, such as degradation tags, are also required to be at the extreme C-terminus. Again, these domains cannot function when internal to a coding region, and are termed Tail domains.

In genetics, a promoter is a region of DNA that initiates transcription of a particular gene. Promoters are located near the genes they transcribe, on the same strand and upstream on the DNA (towards the region of the anti-sense strand, also called template strand and non-coding strand). Promoters can be about 100–1000 base pairs long. For the transcription to take place, the enzyme that synthesizes RNA, known as RNA polymerase, must attach to the DNA near a gene. Promoters contain specific DNA sequences (Jadwiga Bienkowsk, 2002) and response elements that provide a secure initial binding site for RNA polymerase and for proteins called transcription factors that recruit RNA polymerase. These transcription factors have specific activator or repressor sequences of corresponding nucleotides that attach to specific promoters and regulate gene expressions.

Protein structure prediction is the prediction of the three-dimensional structure of a protein from its amino acid sequence  that is, the prediction of its secondary, tertiary, and quaternary structure from its primary structure. Structure prediction is fundamentally different from the inverse problem of protein design. Protein structure prediction is one of the most important goals pursued by bioinformatics and theoretical chemistry; it is highly important in medicine (for example, in drug design) and biotechnology  (for example, in the design of novel enzymes). Every two years, the performance of current methods is assessed in the CASP experiment (Critical Assessment of Techniques for Protein Structure Prediction). A continuous evaluation of protein structure prediction web servers is performed by the community project CAMEO3D (P. Maji, 2004).

Proteins are chains of amino acids joined together by peptide bonds. Many conformations of this chain are possible due to the rotation of the chain about each C$\alpha$ atom (P. Maji, 2004). It is these conformational changes that are responsible for differences in the three dimensional structure of proteins. Each amino acid in the chain is polar, i.e. it has separated positive and negative charged regions with a free C=O group, which can act as hydrogen bond acceptor and an NH group, which can act as hydrogen bond donor. These groups can therefore interact in the protein structure. The 20 amino acids can be classified according to the chemistry of the side chain which also plays an important structural role. Glycine takes on a special position, as it has the smallest side chain, only one Hydrogen atom, and therefore can increase the local flexibility in the protein structure.

## Related Works

Gish et al has proposed database similarity search for identifying protein coding regions, Salzburg et al has proposed a decision tree algorithm to solve the problem. Very less work was done to find the promoter regions in DNA sequences. The Objective of structure prediction is to identify whether the amino acid residue of protein is in helix, strand or any other shape. In 1960 as a initiative step (Debasis Mitra, 2004) of structure prediction the probability of respective structure element is calculated for each amino acid by taking single amino acid properties consideration (P. Maji, **2004**).The third generation technique includes machine learning, knowledge about proteins, several algorithms which gives 70% accuracy. Neural Networks (P. Kiran Sree, 2009).are also useful in implementing structure prediction programs like PHD, SAM-T99.

## Design of MACA based Pattern Classifier

This model is built describing a predefined set of data classes. A sample set from the database, each member belonging to one of the predefined classes, is used to train the model. The training phase is termed as supervised learning of the classifier. Each member may have multiple features. The classifier is trained based on a specific metric. Subsequent to training, the model performs the task of prediction in the testing phase. Prediction of the class of an input sample is done based on some metric, typically distance metric.

The evolution process is directed by the popular Genetic Algorithm (GA) with the underlying philosophy of survival of the fittest gene. This GA framework can be adopted to arrive at the desired CA rule structure appropriate to model a physical system. The goals of GA formulation are to enhance the understanding of the ways CA performs computations and to learn how CA may be evolved to perform a specific computational task and to understand how evolution creates complex global behavior in a locally interconnected system of simple cells. The genetic algorithm is a robust general purpose optimization technique, which evolves a population of solutions

GA is a search technique that has a representation of the problem states and also has a set of operations to move through the search space. The states in the GA are represented using a set of chromosomes. Each chromosome represents a candidate solution to the problem. The set of candidate solutions forms a population. In essence, the GA produces more generations of this population hoping to reach a good solution for the problem. Members (candidate solutions) of the population are improved across generation through a set of operations that GA uses during the search process. GA has three basic operations to expand a candidate solution into other candidate solutions

A CA consists of a number of cells organized in the form of a lattice. It evolves in discrete space and time. The next state of a cell depends on its own state and the states of its neighboring cells. In a 3-neighborhood dependency, the next state $q_i(t + 1)$ of a cell is assumed to be dependent [36] only on itself and on its two neighbors (left and right), and is denoted as

$$q_i(t + 1) = f(q_{i-1}(t), q_i(t), q_{i+1}(t)) \qquad (1)$$

where $q_i(t)$ represents the state of the $i^{th}$ cell at $t^{th}$ instant of time, $f$ is the next state function and referred to as the rule of the automata. The decimal equivalent of the next state function, as introduced by Wolfram, is the rule number of the CA cell. In a 2-state 3-neighborhood CA, there are total 223 that is, 256 distinct next state functions .Out of 256 rules, two rules 85 and 238 are illustrated below:

Rule 85 : $q_i(t + 1) = q_{i+1}(t)$        (2)

Rule 238 : $q_i(t + 1) = q_i(t) + q_{i+1}(t)$     (3)

An n-bit MACA with k-attractor basins can be viewed as a natural classifier. It classifies a given set of patterns into k number of distinct classes, each class containing the set of states in the attractor basin. To enhance the classification accuracy of the machine, most of the works have employed MACA Fig 1, to classify patterns into two classes (say I and II). The following example illustrates an MACA based two class pattern classifier.

## Spatial MACA Tree Building

Input: The Training set S = {S1, S2, · ·, SN}

Output: MACA Tree.

   Partition(S, N)

Step 1: Generate a –spatial MACA with N number of attractor basins.

Step 2: Distribute S based on fitness into N attractor basins (nodes).

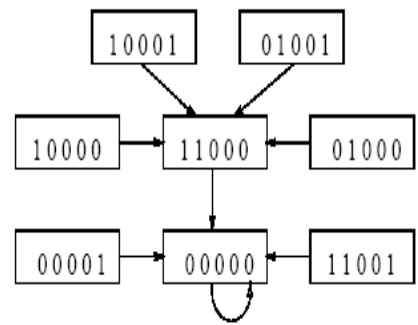

FIGURE 1 EXAMPLE OF MACA WITH BASIN 0000

Step 3: Evaluate the distribution as per rule in each attractor basin

Step 4: If S' of an attractor basin belong to only one class, then label the attractor basin (leaf node) for that spatial class.

Step 5: For examples (S') of an attractor basin belong to N' number of classes, then Partition (S', N').

Step 6: Stop.

A class of non-linear CA, termed as Multiple Attractor CA (MACA) [32], has been proposed to develop the model. Theoretical analysis, reported in this chapter, provides an estimate of the noise accommodating capability of the proposed MACA based associative memory model. Characterization of the basins (P. Maji, 2004). Of attraction of the proposed model establishes the sparse network of non-linear CA ( MACA) (P. Maji, 2004).as a powerful pattern recognizer for memorizing unbiased patterns. It provides an efficient and cost-effective alternative to the dense network of neural net for pattern recognition. Detailed analysis of the MACA rule space establishes the fact that the rule subspace of the pattern recognizing/classifying CA lies at the edge of chaos. Such a CA fig 2, as projected in, is capable of executing complex computation. The analysis and experimental results reported in the current and next chapters confirm this viewpoint. A MACA employing the CA rules at the edge of chaos is capable of performing complex computation associated with pattern recognition.

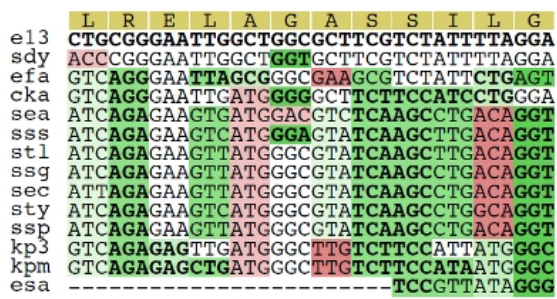

FIGURE 2 EXAMPLE OF SEQUENCES

The entropy and mutual information of the CA in successive generations of GA are reported in Fig 5,6 ,7,8 for four different CA size ($n$= 10, 15, 20, 30). For each of the cases, the values of entropy and mutual information reach their steady state once the AIS FMACA for a given pattern set gets evolved. For understanding the motion, the initial population (IP) is randomly generated. All these figures points to the fact that as the CA evolve towards the desired goal of maximum pattern recognizing capability, the entropy values fluctuate in the intermediate generations, but saturate to a particular value (close to the critical value 0·84 when fit rule is obtained. Simultaneously, the values of mutual information fluctuate at the intermediate points prior to reaching maximum value that remains stable in subsequent generations. All these figures indicate that the CA move from chaotic region to the edge of chaos to perform complex computation associated with pattern recognition.

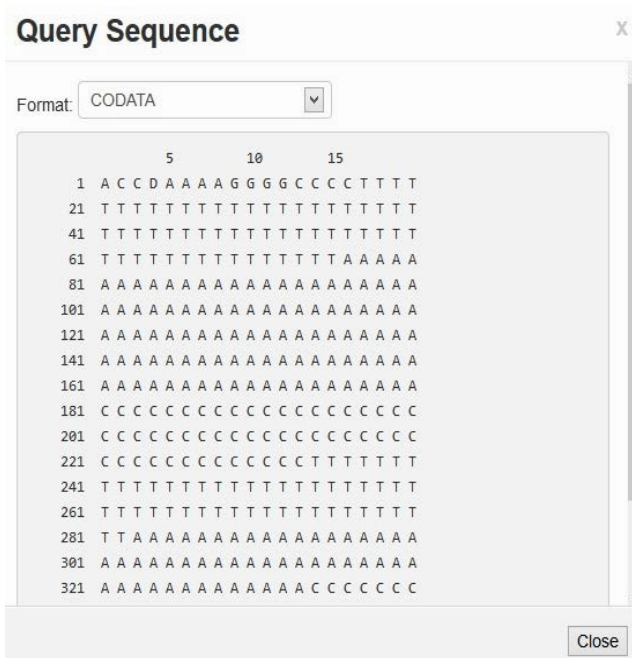

FIGURE 3 AMINO ACID SEQUENCE

Crossover operator randomly chooses a locus and exchanges the subsequences before and after that locus between two chromosomes to create two offspring. The crossover operator helps to explore the search space by virtue of providing means to generate new solutions out of the current population. The probability of selecting a pair of chromosomes to be employed for crossover depends upon the fitness of chromosomes. Majority of chromosomes for the next generations are produced through the crossover process. We have experimented with two different techniques of crossover to speed up the rate of convergence. However, there is no major difference in performance between the two schemes.

Mutation operator randomly flips some of the bits in a chromosome. The mutation, applied to perturb one or more solutions, ensure that the search space explored is not closed under crossover. The probability of mutation on a given chromosome is kept very low. In our problem, 10% of the population of NP is generated out of mutations of the elite rules. Here also we have experimented with two standard techniques of mutation.

### Experimental Step

- Experiments were conducted with wide range of data sets.

- We have used ENCODE, BG570, HMR195, Fickett and Tongue, ASP67 data sets for evaluating the proposed classifier.

- Using Bp, create an input sequences Ib (corresponding to the base CA protein) by replacing each amino acid in the primary structure with its hydrophobia city value. The output sequences Ob is created by replacing the structural elements in Bs with the values, 200, 600, 800 for helix C, strand and coil respectively

- Solve the system identification problem, by performing CA de convolution with the output sequences Ob and the input sequence Ib to obtain the CA response, or the sought after running the algorithm.

- Transform the amino acid sequence of Tp into a discrete time sequences It, and convolve with F; thereby producing the predicted structure (Ot = It*F) of the target CA protein Fig 3, 4, 5.

- The result of this calculation Ot is a vector of numerical values. For values between 0 and 200, a helix C is predicted, and between 600

and 800, a strand is predicted by CA. All other values will be predicted as a coil by MACA. This produces mapping for the required target structure Ts of the target CA protein T

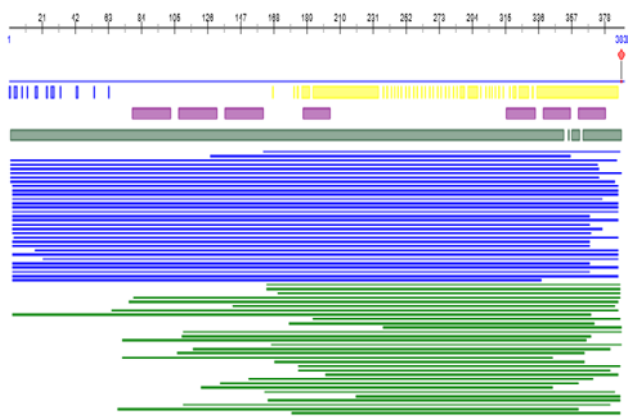

FIGURE 4 PROTEIN STRUCTURE PREDICTION INTERFACE WITH GREEN AS HELIX AND BLUE AS PROTEIN CODING

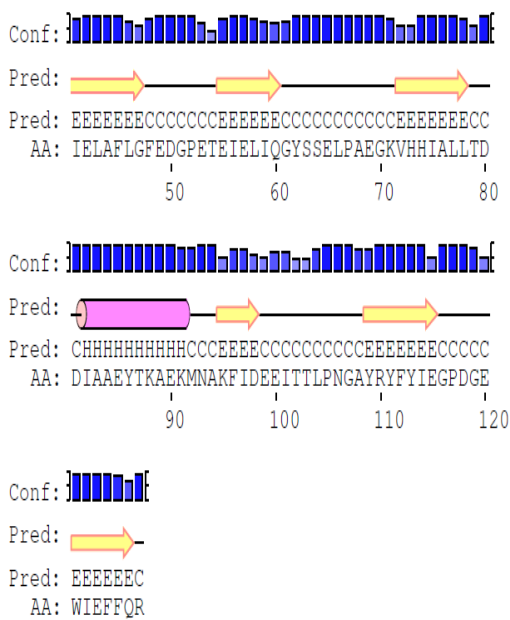

FIGURE 5 PROTEIN STRUCTURE AND PROMOTER PREDICTION ANALYSIS

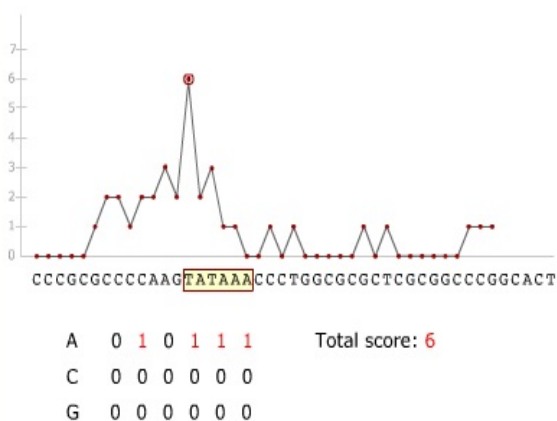

FIGURE 6 PROMOTER PREDICTION ANALYSES

## Experimental Results

In the experiments conducted, the base proteins are assigned the values 300,700,900 for helix C, strand and coil respectively. We have found an structure numbering scheme that is build on Boolean characters of CA which predicts the coils, stands and helices separately .The MACA based prediction procedure(P. Maji, 2004) as described in the previous section is then executed, and each occurrence of each sequences in the resulting output, is predicted . The query sequence analyzer was designed and identification of the green terminals of the protein is simulated in the figure 4. The analysis of the sequence and the place of joining of the proteins are also pointed out in the figure 5. Experimental results Figure 7, 8 which include the similarity and accuracy tables with each of the components are separately represented.

| Prediction Method | Protein Coding Region | Promoter Identification | Protein Structure Prediction |
|---|---|---|---|
| DSP | 92% | 70% | 96% |
| PHD | 70% | 68% | 84% |
| SAM-T99 | 68% | 77% | 87% |
| SS Pro | 70% | 73% | 81% |
| MACA | 90% | 85% | 97% |

FIGURE 7 PREDICTION ACCURACY OF PROTEIN CODING REIGNS AND PROMOTER REGIONS

| Target: 1PFC | Prediction Accuracy | Target: 1PP2 | Prediction Accuracy | Target: 1QL8 | Promoter Accuracy |
|---|---|---|---|---|---|
| Exp 1 | 64% | Exp 5 | 75% | Exp 9 | 85% |
| Exp 2 | 66% | Exp 6 | 90% | Exp 10 | 90% |
| Exp 3 | 69% | Exp 7 | 83% | Exp 11 | 82% |
| Exp 4 | 71% | Exp 8 | 87% | Exp 12 | 91% |

FIGURE 8 PREDICTION ACCURACY FOR PROTEIN STRUCTURE PREDICTION, PROMOTER IDENTIFICATION AND PROTEIN CODING REGION IDENTIFICATION

## Conclusion

We have proposed a cellular automata classifier which can address major issues in bioinformatics. The proposed tool was tested with datasets of different lengths. Extensive experiments were conducted for reporting the accuracy of the proposed tool. The average accuracy of MACA when tested with ENCODE, BG570, HMR195, Fickett and Tongue, ASP67 datasets is 78%. At the end of all experiments MACA accuracy for prediction protein coding regions is 94%. This work can be extended to achieve good classification accuracy for other problems in bioinformatics like genome annotation, sequence analysis etc. So with good heuristic fitness equations and values we can achieve more than 84% average accuracy with MACA for all these addressed problems.

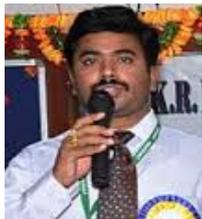

**Prof P. Kiran Sree** is working as Professor in department of CSE in BVC Engineering College. He has published forty technical papers in international journals and conferences. His areas of interests include parallel algorithms, artificial intelligence, and compiler design and computer networks. He was the reviewer for many IEEE Society conferences and Journals in artificial intelligence and networks. His bibliography is listed in Marquis Who's Who in the World, 29th Edition (2012), USA.

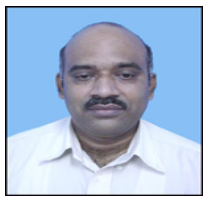

**Dr Inampudi Ramesh Babu** is working as a Professor in the Department of Computer Science& Engineering, Acharya Nagarjuna University. Also, he has been an Academic Senate member of the same university since 2006. He held many positions in the Acharya Nagurjuna University as Head, Director of the Computer Centre, Chairman of the PG Board of studies, Member of the Executive Council, Special Officer, Additional Convener of ICET Examinations and Convener of MCA Admissions. He is currently supervising ten PhD students who are working in different areas of image processing and artificial intelligence. He has published 100 papers in international journals and conferences.

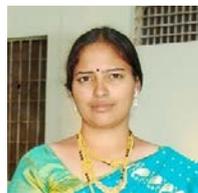

**Smt S. S. S. N. Usha Devi. N** is working as Assistant Professor in department of CSE in University College of Engineering, JNTU Kakinada. She has published five papers in international journals and four in international conferences. She is the member of IAENG.